\documentclass[a4paper]{aa}
\usepackage{txfonts,graphicx,natbib}
\bibpunct{(}{)}{;}{a}{}{,}

\newcommand{\pq}{\ensuremath{P_Q}}
\newcommand{\pu}{\ensuremath{P_U}}

\begin{document}
\title{Polarimetry of the dwarf planet (136199) Eris
       \thanks{Based on observations made with ESO Telescopes at the
               Paranal Observatory under programme ID 178.C-0036
               (PI: A.\ Barucci)}}
       \author{
        I.~Belskaya     \inst{1}
       \and
        S.~Bagnulo      \inst{2}
       \and
        K.~Muinonen     \inst{3}
       \and
        M.A.~Barucci    \inst{4}
       \and
        G.P.~Tozzi      \inst{5}
       \and
        S.~Fornasier    \inst{4,6}
       \and
        L.~Kolokolova   \inst{7}
        }
\institute{Astronomical Observatory of Kharkiv National University,
           35 Sumska str., 61022 Kharkiv, Ukraine.\\
           \email{irina@astron.kharkov.ua}
           \and
           Armagh Observatory,
           College Hill,
           Armagh BT61 9DG,
           Northern Ireland, U.K.
           \email{sba@arm.ac.uk}
           \and
           Observatory, PO Box 14, 00014 University of Helsinki, Finland.
           \email{muinonen@cc.helsinki.fi}
           \and
           LESIA, Observatoire de Paris, 5,
           pl.~Jules Janssen, FR-92195 Meudon cedex, France.\\
           \email{antonella.barucci@obspm.fr; sonia.fornasier@obspm.fr}
           \and
           INAF - Oss. Astrofisico di Arcetri,
           Largo E. Fermi 5, I-50125 Firenze, Italy.
           \email{tozzi@arcetri.astro.it}
           \and
           University of Paris 7  "Denis Diderot",  10 rue Alice Domon et
           Leonie Duquet, 75013 Paris, France.
           \and
           University of Maryland, College Park, MD, USA.
           \email{ludmilla@astro.umd.edu}
}
\authorrunning{I.\ Belskaya, S.\ Bagnulo, K.\ Muinonen, et al.}
\titlerunning{Polarimetry of the dwarf planet Eris}

\date{Received: July 7 2007 / Accepted: 26 November 2007}
\abstract
{Study the surface properties of the transneptunian population of
  Solar-system bodies}
{ We investigate the surface characteristics of the large dwarf planet
  (136199)\,Eris.  }
{ With the FORS1 instrument of the ESO VLT, we have obtained Bessell
  broadband $R$ linear polarimetry and broadband $V$ and $I$
  photometry of Eris. We have modelled the observations in terms of
  the coherent-backscattering mechanism to constrain the surface
  properties of the object.}
{ Polarimetric observations of Eris show a small negative linear
  polarization without opposition surge in the phase angle range of
  $0.15-0.5\degr$.  The photometric data allow us to suppose a
  brightness opposition peak at phase angles below $0.2-0.3\degr$. The
  data obtained suggest possible similarity to the polarimetric and
  photometric phase curves of Pluto. The measured absolute magnitude
  and broadband colors of Eris are $H_V$=-1.15 mag, $V-R=0.41$ mag,
  and $V-I=0.75$ mag.}
{The observational data with theoretical modelling are in agreement
  with the surface of Eris being covered by rather large inhomogeneous
  particles.}


\maketitle
\section{Introduction}
The discovery of the Pluto-sized transneptunian object 2003~UB$_{313}$
\citep{Broetal05}, later named (136199)~Eris, was succeeded by an
intense physical characterization phase and no less than a discussion
on the definition of a planet.  After the resolution of IAU (24 August
2006) Eris is classified, together with Pluto, as a member of the new
category of ``the dwarf planets''. Eris was discovered near its
aphelion at 97\,AU from the Sun and has an orbit with high
eccentricity (0.44) and inclination ($44\degr$). According to its
dynamical characteristics, it belongs to the so-called ``detached
objects'' with pericenters decoupled from Neptune \citep{Glaetal07}.

The first size estimation of Eris was carried out by \citet{Beretal06}
based on thermal emission, resulting in a diameter of $3000 \pm
400$\,km and a geometric albedo of $0.60 \pm 0.15$. Later on,
\citet{Broetal06}, using the Hubble Space Telescope, measured directly
the angular size of Eris corresponding to the diameter of $2400 \pm
100$\,km, about 5\,\% larger than that of Pluto, and derived the
albedo of $0.87 \pm 0.05$ in the $V$ band. The discrepancy in the
albedo is rather large but, undoubtedly, Eris is a high-albedo object
like Pluto \citep[see][]{Buretal03}. According to \citet{BroSchal07},
the mass of Eris exceeds the mass of Pluto by a factor of 1.27.

The spectral properties of Eris are also found to be rather similar to
those of Pluto. The near-infrared spectrum is dominated by methane ice
absorptions \citep{Broetal05}. However, the central wavelengths of the
CH$_{4}$ bands do not show the shift in wavelength seen in the spectra
of Pluto and Triton which is attributed to the fact that most (Triton)
or some (Pluto) of the CH$_{4}$ is dissolved in solid N$_{2}$. The
quality of Eris' spectrum does not allow the detection of
N$_{2}$. \citet{Licetal06} reported a visible spectrum of Eris in
which the CH$_{4}$ ice bands show a small shift in the band at around
0.89\,$\mu$m and they suggested that the shift could indicate the
presence of methane diluted in N$_{2}$. \citet{Dumetal07}, using the
new SINFONI instrument installed at the ESO VLT (Paranal Observatory,
Chile), obtained a high-quality near-infrared spectrum and, on the
basis of modelling, they obtained the best
fit of the spectrum including N$_{2}$ on the surface of Eris.
Due to the highly eccentric orbit, they expect a thin
atmosphere close to perihelion, which freezes out onto the surface
when the object is located at a larger heliocentric distance. They
modelled the surface using two types of terrains of distinct
composition: about $50\,\%$ covered with pure methane ice, while the
other part is made of an intimate mixture of methane, nitrogen, and
water and tholin ices. The best model fits are obtained when the icy
grains are rather large, from sub-mm to a few tens of mm in size
\citep{Dumetal07}.

In the present paper, we report the first polarimetric observations of
Eris made as part of the Large-Program observations at ESO VLT,
aimed to set constraints on its surface properties.

\begin{table*}
\caption{Polarimetry of (136199)\,Eris in the Bessell $R$ band. $\pq$
and $\pu$ are the transformed Stokes parameters in a way that $\pq$
represents the flux perpendicular to the plane Sun-Object-Earth (the
scattering plane) minus the flux parallel to that plane, divided by
the sum of the two fluxes, and $\Theta$ is the position angle
measured counterclockwise from the perpendicular to the scattering
plane.} \label{Tab_Pol} \centering
\begin{tabular}{ccccrrr}
\hline \hline
Date                             &
Time (UT)                        &
Sky                              &
Phase angle                      &
\multicolumn{1}{c}{$\pq$}        &
\multicolumn{1}{c}{$\pu$}        &
\multicolumn{1}{c}{$\Theta$}     \\
\multicolumn{1}{c}{(yyyy mm dd)} &
\multicolumn{1}{c}{(hh:mm)}&
\multicolumn{1}{c}{}        &
\multicolumn{1}{c}{(DEG)} &
\multicolumn{1}{c}{(\%)} &
\multicolumn{1}{c}{(\%)}         &
\multicolumn{1}{c}{(DEG)}        \\
\hline
2006 10 19 & 03:27 & PHO & 0.15 & $-0.09 \pm 0.07$ & $ 0.00 \pm 0.07$ & $ 91 \pm 23$ \\
2006 11 12 & 02:21 & CLR & 0.30 & $-0.06 \pm 0.07$ & $-0.12 \pm 0.07$ & $122 \pm 19$ \\
2006 11 18 & 02:25 & THN & 0.35 & $-0.32 \pm 0.12$ & $ 0.00 \pm 0.10$ & $ 90 \pm  9$ \\
2006 12 14 & 01:53 & PHO & 0.51 & $-0.12 \pm 0.07$ & $-0.05 \pm 0.07$ & $100 \pm 21$ \\
\hline
\end{tabular}
\end{table*}
\begin{table}
\caption{Photometry of (136199)\,Eris. The errors on $R$ magnitudes
are estimated \textit{a priori} (see text).}\label{Tab_Phot}
\centering
\begin{tabular}{cccccr}
\hline \hline
Date                &
Time (UT)  &
Phase angle&
Band       &
Magnitude  &\\
\multicolumn{1}{c}{ (yyyy mm dd)} & \multicolumn{1}{c}{(hh:mm)} &
\multicolumn{1}{c}{(DEG)}
                                 &
                                 &
                                 &
                                  \\
\hline
2006 10 19 & 03:27 & 0.15 &$R$&$18.34 \pm 0.05$ \\
2006 10 20 & 05:42 & 0.16 &$V$&$18.74 \pm 0.02$ \\
2006 10 20 & 05:45 & 0.16 &$I$&$18.00 \pm 0.02$ \\
2006 10 20 & 05:49 & 0.16 &$V$&$18.75 \pm 0.02$ \\
2006 10 20 & 05:52 & 0.16 &$I$&$17.99 \pm 0.02$ \\
2006 11 12 & 02:21 & 0.30 &$R$&$18.37 \pm 0.05$ \\
2006 12 14 & 01:53 & 0.51 &$R$&$18.40 \pm 0.05$ \\
\hline
\end{tabular}
\end{table}
\section{Observations and data reduction}
Observations of Eris have been obtained at the ESO VLT with the
FORS1 instrument in service mode during an observing period from
October to December 2006.

FORS1 is a multi-mode instrument for imaging and spectroscopy equipped
with polarimetric optics. For the present study, FORS1 has been used
to measure the broadband linear polarization of Eris at four
different epochs in the Bessell $R$ filter. Besides, additional
photometric observations in the $V$ and $I$ filters were made.

Taking advantage of the flexibility offered by the VLT service
observing mode, we distributed the observations along three months to
obtain data points in the maximum possible range of phase angles for
the ground-based observations, setting precise time intervals for the
execution of the observations. In order to reduce the impact of
sky background on the data quality, we generally tried to avoid
observations with the target close to the Moon, and with a large
fraction of lunar illumination.

Polarimetric observations were performed with the retarder waveplate
at all positions between 0 and 157.5\degr\ (at 22.5\degr\ steps),
setting the exposure time of each frame to 180\,s. The frames were
then processed and combined as explained in \cite{Bagetal06}. In
addition to the data reduction procedure described in \cite{Bagetal06}, the
observed position angle was rotated for Eris by -1.23\degr\ to
compensate for a small chromaticity effect of the instrument retarder
waveplate. Further details can be found in the FORS1 user manual and
in \citet{Fosetal07}.

The errors were calculated taking into account both photon noise and
background subtraction. We found that, for apertures of about 7--8
pixels, the error due to the photon noise was similar to the error
introduced by the background subtraction (for smaller apertures, the
former is the dominating source of error, and vice versa). From the
observations obtained on 2006-10-19, 2006-11-12, and 2006-12-14, we
measured a polarization generally independent of the aperture.
For these observations, we finally adopted the polarization measured
for apertures of 6 to 8 pixels (note that, for such small aperture
values, the fluxes measured in the ordinary and extra-ordinary beams
are still increasing with the aperture).  By contrast, from the
observations taken on 2006-11-18 we found that the polarization
strongly depends on the aperture, varying from about 0.2\,\% for a 5
pixel aperture, up to about 0.6\,\% for an aperture of 20 pixels. We
found that by discarding the frame obtained at 22.5\degr, the
polarization is more independent of the aperture, varying from about
0.1\,\% for a 5 pixel aperture, to about 0.4\,\% for a 20 pixel
aperture.  However, we could not find something obviously wrong with
the raw data obtained with the retarder waveplate at 22.5\degr.  For
the point obtained on 2006-11-18, we finally adopted a value of about
0.3\,\%, which is consistent with the results obtained by discarding
the frame obtained at 22.5\degr.

The results of polarimetric observations are given in Table 1, which
contains the epoch of the observations (date and UT time), the
night-time sky conditions (THN = thin cirrus, CLR = clear, PHO =
photometric), the phase angle, the measured Stokes parameters \pq and
\pu, and the position angle of the linear polarization $\Theta$,
transformed relative to the scattering plane as explained in
\citet{Bagetal06} and \citet{Lanetal07}.

The linear polarization of the light scattered by the surface of Eris
in the phase-angle range of $0.15-0.51\degr$ is found to be
small, never exceeding a 3\,$\sigma$ detection.

Acquisition images were used to obtain photometry, although not all
observing nights were clear. In particular, bad sky transparency
during the night of 2006-11-18 did not allow us to obtain a proper
estimate of the magnitude. We associated an error of 0.05~mag to the
photometry, which is consistent with the zero points of the
calibration plan, with the errors of photon noise and background being
negligible in comparison with those of the zero points. Additional
photometric observations in the $V$ and $I$ bands were obtained during
the photometric night of 2006-10-20 at the phase angle of 0.16 deg
close to the minimum phase angle reachable for Eris in this
apparition. The integration times in the $V$ and $I$ bands were 100\,s
and 200\,s, respectively, in order to obtain high-signal-to-noise-ratio
images. The images were taken when the object was at the airmass of
1.13. The data were reduced in a standard manner described in details
by \citet{Peix04}. Results of photometric measurements are given in
Table 2. They allowed for the estimation of the absolute magnitude and
broadband colors of Eris: $H_V$=-1.15 mag, $V-R=0.41$ mag, and
$V-I=0.75$ mag.

\section{Results and discussion}
\begin{figure}
\includegraphics*[width=10cm,angle=0]{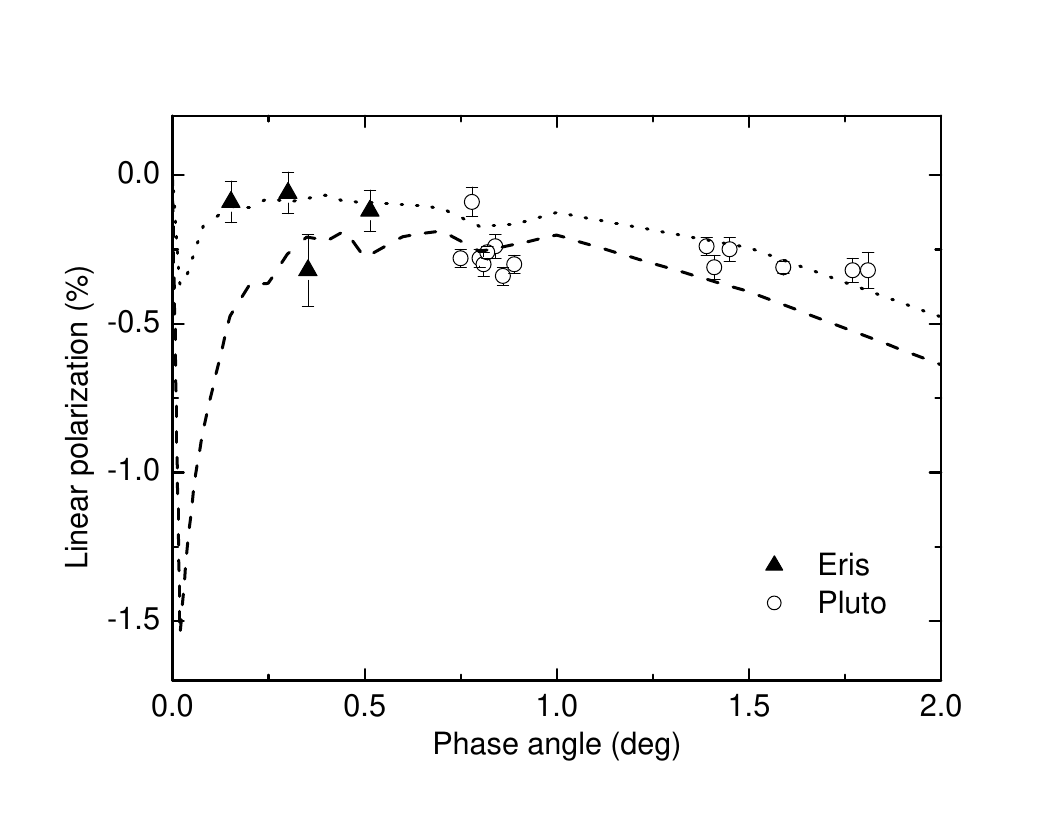}
\caption{\label{Fig_Polar} Linear polarization observations vs. phase angle
for Eris and Pluto (for references, see text). The dashed curves are the
results of our modelling and show the envelope of the polarization
curves for the given range of geometric albedos (see Sect.~3).
The Eris 0.35\degr\ point was not included in the derivation of the envelope.}
\end{figure}

Polarimetry is a powerful tool to investigate the physical properties
of solid surfaces of Solar-system bodies. At small phase angles, these
objects typically exhibit a phenomenon of \textit{negative
  polarization}. It is a peculiar case of partially linearly polarized
scattered light where the electric field vector component parallel
to the scattering plane predominates over the perpendicular
component. Negative linear polarization was first discovered by
\citet{Lyot29} for the Moon and later found to be a ubiquitous
phenomenon for planetary surfaces at small phase angles.  The
linear polarization varies with the phase angle, characterizing
the properties of the surface, such as particle size,
heterogeneity, complex refractive index, porosity, and surface
roughness. The first polarimetric observations for a TNO (except
Pluto) carried out by \citet{Boeetal04} for (28978) Ixion revealed
a pronounced negative polarization noticeably changing as a
function of the phase angle in spite of the small range of phase
angles covered ($0.2\degr-1.3\degr$).  Later, three other
objects have been observed: (2060) Chiron and (50000) Quaoar
\citep{Bagetal06} and 29981 (1999~TD$_{10}$) \citep{Rouetal05}. The
polarimetric data available for TNOs show noticeable negative
polarization with two different trends at small phase angles:
subtle changes in the negative polarization are observed for the
largest objects Pluto and Quaoar, whereas sudden enhancements are
observed for the others. Our observations of Eris do not show
noticeable changes of linear polarization in the phase angle range
covered. Figure~\ref{Fig_Polar} shows the polarimetric
observations of Eris in the $R$ band and Pluto in the visible
light versus phase angle. Note that the observations of Pluto
refer to the Pluto-Charon system and show notable similarity in
the degree of polarization measured in different oppositions by
different authors \citep{KelFix73,BreCoc82,Avretal92}.
Unfortunately, the observations of Eris and Pluto cover different
ranges in the phase angle. However, the objects can be anticipated
to be similar in their surface characteristics.

\begin{figure}
\includegraphics*[width=10cm,angle=0]{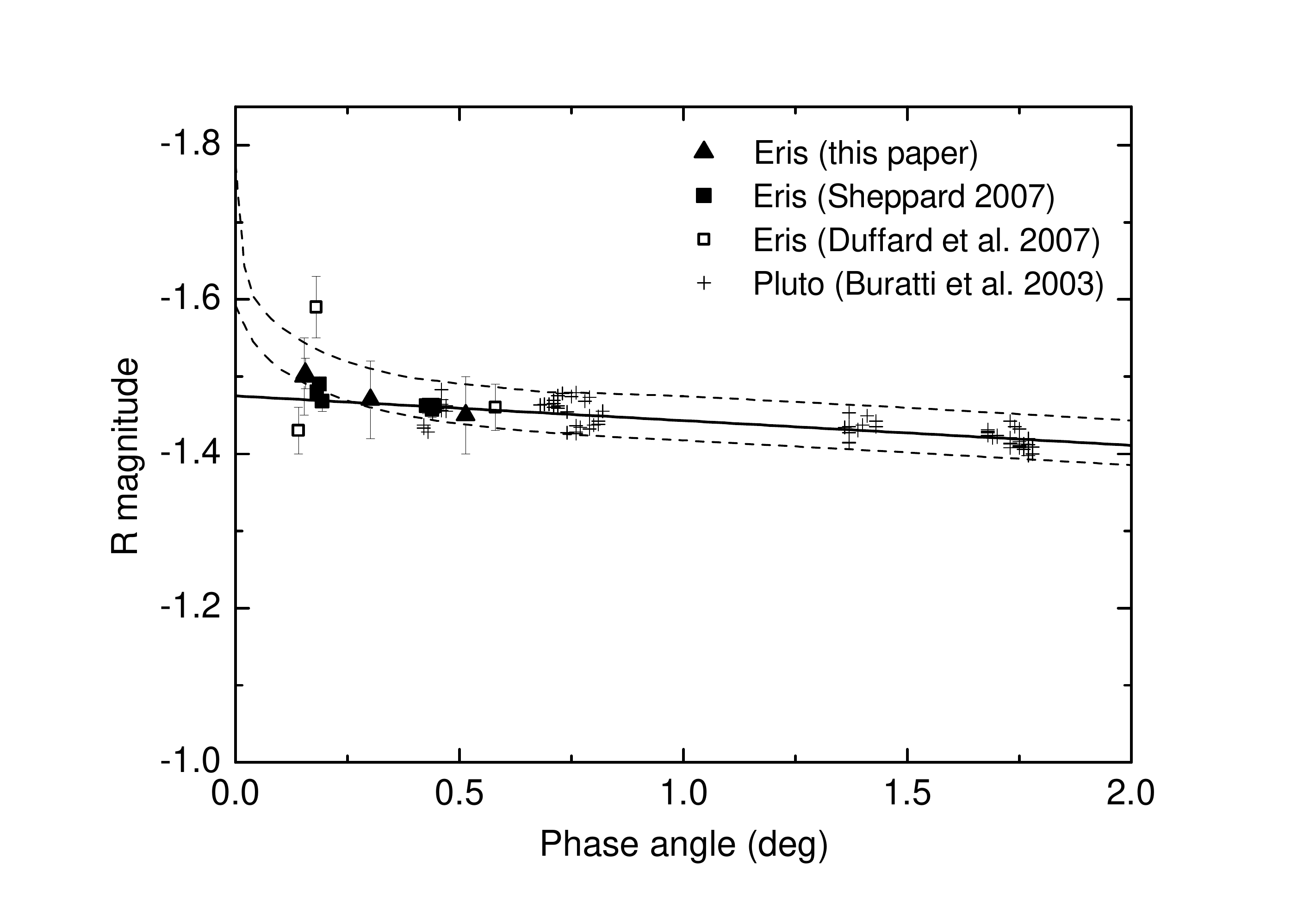}
\caption{\label{Fig_Magni} Reduced $R$ magnitude vs. phase angle
for Eris. Data of Pluto in the $R$ band is plotted according to \citet{Buretal03}
 with an arbitrary shift to Eris data. The solid line shows a linear fit to
all data. The dashed lines show the envelope of the photometric model
curves derived simultaneously with the polarimetric envelope.}
\end{figure}

The small linear polarization near opposition for a bright object like
Eris is somewhat unexpected. According to the mechanism of coherent
backscattering, which is considered as the most probable cause of
negative polarization at small phase angles \citep[for reviews,
  see][]{Muietal02,Belskal07}, a sharp asymmetric surge of negative
polarization is expected for high-albedo objects. \citet{Rosetal97}
and \citet{Rosetal05} report detections of such surges for bright
satellites and asteroids. In all cases, the surges are accompanied by
narrow brightness opposition peaks in an agreement with
coherent-backscattering modelling \citep[for a review, see][]{Misetal06}.

Figure~\ref{Fig_Magni} shows the magnitude phase dependence in the $R$
band for Eris in comparison with Pluto's observations in the same band
plotted according to \citet{Buretal03}. The magnitude phase dependence
for Eris was composed of our observations (see Table~\ref{Tab_Phot}), those from
\citet{Shep07} in 2005 averaged for each observing night, and those
from \citet{Duff07} in 2005-2007 averaged for each observing run. All
observations agree within their uncertainties. The scatter of the data
is probably connected with rotational brightness variations of Eris
which are impossible to take into account since the rotation period
remains unknown.  According to \citet{Duff07} a short-term variability
is very low but a very long rotation period cannot be ruled out. The
data available shows that a long-term variability of Eris (if it
exists) should be less than 0.1 mag contrary to Pluto with a
lightcurve amplitude of 0.33 mag and a rotation period of 6.387 days
\citep{Buietal97}. The data for Pluto in Fig~\ref{Fig_Magni} were shifted with
respect to the data for Eris in order to make the phase curves
coincide in the overlapping phase-angle range. As in the case of
polarimetric phase dependence (Fig.~\ref{Fig_Polar}), both objects show rather
similar trends with phase angle. The linear phase coefficient derived
for Pluto in the phase angle range of $0.4-1.8\degr$ is $0.032 \pm
0.001$ mag/deg \citep{Buretal03}. A linear fit to the Eris data in
the phase-angle range of $0.2-0.6\degr$ gave a phase coefficient of
$0.08 \pm 0.03$ mag/deg.  \citet{Rabetal07} reported a similar
small phase coefficient of $0.10 \pm 0.02$ mag/deg in the $V$
band. However all these estimations are rather uncertain due to large
scatter of data and small range of the covered phase angles. In fact,
data of Pluto and Eris can be fitted by a single linear dependence
with Pluto's phase coefficient (see solid line in Fig~\ref{Fig_Magni}). A small
increase in brightness toward opposition found for Eris is unusual for
bright surfaces typically showing a pronounced opposition surge with a
phase slope of 0.2-0.3 mag/deg in the same phase range of
$0.2-0.6\degr$ \citep[e.g.][]{Veretal05}. An opposition surge can
exist at smaller phase angles not covered by the observations.

According to the data available, the photopolarimetric properties of
Eris and Pluto are expected to be quite similar. When interpreting the
data, we need to take into account the existence of a thin atmosphere
of Pluto and possible thin atmosphere of Eris. As compared to the
satellites of the major planets having atmospheres, the absence of
negative polarization at small phase angles has been reported for
Titan only (Veverka 1973) while, for Io and Europa, the negative
polarization peak was found with an amplitude of about $0.4\,\%$ at
phase angles of $0.2-0.7 \degr$ \citep{RosKis05}. The absence of such
a peak for Eris puts constraints on its surface properties.

Another point which needs to be mentioned is that both objects have
satellites. While the satellite of Eris is small \citep{Broetal06}
and should not affect the observations made, all data discussed above
for Pluto refer to the Pluto-Charon system. \citet{Buietal97} made
an attempt to separate the photometric properties of these objects and
obtained a smaller phase coefficient for Pluto (0.029 mag/deg) as
compared to the Pluto-Charon system (0.032 mag/deg). The contribution
of Charon to the linear polarization observed for the Pluto-Charon system
remains unknown.

Below we present results of modelling of the photopolarimetric
properties of Eris and the Pluto-Charon system assuming similar
behaviors for their photometric and polarimetric phase curves.

We explain the polarimetric phase curves of Eris and Pluto through
numerical simulations of coherent backscattering by Rayleigh
scatterers in spherical media \citep{Muinonen04}, assuming a model
composed of two kinds of scattering media \citep[e.g.,][]{Boeetal04,
Bagetal06}. There are five parameters in the model: two
single-scattering albedos, two mean free paths, and the weight
factor $w_{\rm d}$ for the darker component (the weight factor of
the brighter component being $w_{\rm b}=1-w_{\rm d}$).

We make use of existing computations for 630 spherical random media:
the data base entails 21 single-scattering albedos
$\tilde{\omega}=0.05$, 0.10, \ldots, 0.90, 0.95, 0.97, 0.99, and 30
dimensionless mean free paths $k\ell = 2\pi \ell/\lambda = 10$, 20,
30, \ldots, 100, 120, 140, \ldots, 200, 250, 300, \ldots, 400, 500,
600, \ldots, 1000, 2000, 3000, \ldots, 5000, 10000 ($k$ and $\lambda$
are the wave number and wavelength). The spherical media have, in
essence, infinite diametrical optical thicknesses, mimicking
macroscopic objects.

Figure~\ref{Fig_Polar} shows the envelope of the extreme values of
polarization curves when fixing the model geometric albedo at $p_R =
0.600$, 0.601, 0.602, $\ldots$ , 0.900, and, for each albedo, updating
the polarization envelope at each phase angle if the rms is less than
0.12 (roughly twice the observational error). Note that the boundaries
of the envelope do not correspond to any single model polarization
curve but rather outline the regime of plausible curves. The Eris
observation at the phase angle of 0.35\degr\ was omitted as an
outlier. Some of the polarimetric observations of Pluto are
located on or outside the boundaries of the envelope computed. We
consider this systematic feature to be mainly due to the limitations
of the two-component Rayleigh-scatterer model utilized.

A photometric opposition-effect envelope was derived in a way
resembling the derivation of the polarimetric envelope. The
photometric envelope is defined, at a given phase angle, by the
maximum and minimum ratios of disk-integrated brightness at that phase
angle and at exact opposition. In order to obtain reasonable fits, the
coherent-backscattering peaks were multiplied by an additional
linear-exponential function with a peak width and amplitude of
0.2\degr\ and 10\,\%, respectively
\citep[cf.][]{Rouetal05}. This additional enhancement can be
envisaged to rise from shadowing contributions. Alternatively, the
need for the additional function could be removed through more
sophisticated coherent-backscattering modeling.  In
Fig.~\ref{Fig_Magni}, the photometric envelopes are depicted after
introducing a small separation into the envelope curves for better
illustration.

Based on the envelopes above and the opposition surge suggested in
linear polarization and brightness phase curves at extremely small phase
angles, we can conclude that the mean free paths in the regoliths of
Eris and Pluto are most probably long, potentially so long that the
related opposition phenomena cannot be observed any more. For
realistic high-albedo regolith structures, long mean free paths
suggest that the scattering centers are located far from each
other. One of the explanations of this finding may be that the
regolith particles are rather transparent large particles with
internal or surface inhomogeneities responsible for the
coherent-backscattering phenomena. Such an explanation is in agreement
with the spectroscopic findings reviewed above.

\section{Conclusions}\label{Sect_Discussion}
The present polarimetric observations of Eris constitute the first
observations of a high-albedo transneptunian object at very small
phase angles. A sharp surge of negative polarization expected for
high-albedo objects within the mechanism of coherent backscattering
was not found in the measured range of phase angles
($0.15-0.5\degr$). It was also shown that, if a brightness opposition
peak exists for Eris, it must be constrained to phase angles below
$0.2-0.3\degr$. The data obtained for Eris suggest possible similarity
with the polarimetric and photometric phase curves of Pluto. Their
modelling puts constraints on the surface properties suggesting that
the regoliths of Eris and Pluto most probably consist of transparent
inhomogeneous particles large compared to the wavelengths of visible
light.

\end{document}